\documentclass[twocolumn,prl,showpacs,floatfix]{revtex4-1}
\usepackage{graphicx}
\usepackage{epsfig}
\usepackage{rotating}
\usepackage{amsmath}
\usepackage{amsfonts}
\usepackage{amssymb}
\usepackage{enumerate}
\usepackage{longtable}
\usepackage{dcolumn}% Align table columns on decimal point
\usepackage{bm}
\usepackage{physics}
\usepackage{verbatim}

\usepackage{amsmath,amsthm,amsfonts,amssymb}

\usepackage{color}
\numberwithin{equation}{section}
\numberwithin{figure}{section}

\DeclareMathAlphabet{\mathpzc}{OT1}{pzc}{m}{it}

\newcommand{\be}{\begin{equation}}
\newcommand{\ee}{\end{equation}}

\newcommand{\cZ}{{\cal Z}}

\begin{document}

\title{Nonzero temperature Entanglement Negativity of quantum spin models: Area law, Linked Cluster Expansions and Sudden Death}
\author{Nicholas E. Sherman$^1$, Trithep Devakul$^2$, Matthew B. Hastings$^3$ and Rajiv R. P. Singh$^1$}
\affiliation{$^1$Department of Physics, University of California Davis, CA 95616, USA \\ $^2$Department of Physics, Princeton University, NJ 08544, USA 
\\ $^3$Quantum Architectures and Computation Group, Microsoft Research, Redmond, WA 98052, USA}

\date{\rm\today}

\begin{abstract}
We show that the bipartite logarithmic entanglement negativity (EN) of quantum spin models obeys an area law at all nonzero temperatures.
We develop numerical linked cluster (NLC) expansions for the `area-law' logarithmic entanglement negativity 
as a function of temperature and other parameters. For one-dimensional models the results of NLC are compared with exact
diagonalization on finite systems and are found to agree very well. The NLC results are also obtained for two dimensional
XXZ and transverse-field Ising models.
In all cases, we find a sudden onset (or sudden death) of negativity at a finite temperature above which the negativity is zero.
We use perturbation theory to develop a physical picture for this sudden onset (or sudden death).
The onset of EN or its magnitude are insensitive to classical finite-temperature phase
transitions, supporting the argument for absence of any role of quantum mechanics at such transitions.
On approach to a quantum critical point at $T=0$, negativity shows critical scaling in size and temperature.
\end{abstract}

%\pacs{74.70.-b,75.10.Jm,75.40.Gb,75.30.Ds}

\maketitle
\section{Introduction}

In recent years, the study of quantum entanglement in many-body systems has been a topic of great interest \cite{cardy,rmp-review,rmp2,rmp3,cft1,huerta,max,peschel,cardy-rev,grover14,wolf08,melko,series-prl,oitmaa,nlc,nlc-dmrg,roscilde,wessel,devakul,inglis}.  
It sheds light on many fundamental issues in quantum statistical mechanics from quantum phase transitions and universality to thermalization and many-body localization.  
In studying bipartite entanglement in pure states, the Von-Neumann entropy provides the ideal measure and together with its easier to compute generalization
of the Renyi entropies gives a rather complete quantitative description of entanglement in such systems. In contrast, for a mixed state, these entropy based measures 
are often dominated by classical probabilities and isolating quantum entanglement becomes non-trivial. Even bipartite Mutual Information (MI), which is a measure 
of all correlations between subsystems, at nonzero temperatures, is usually dominated by classical correlations and not quantum entanglement.

The notion of separability provides a concrete definition for the absence of entanglement. If a mixed state density matrix can be decomposed
as a sum of pure state density matrices, each of which is separable between subsystems A and B and has zero entanglement entropy, one can clearly conclude that there is
no entanglement between A and B. However, examining separability requires
trying all possible decompositions of a mixed state in terms of pure states \cite{rmp-review,rmp2,rmp3}, which is
not easy, nor does it provide a quantitative measure of how entangled a non-separable system is. 
Many measures of entanglement in a mixed state have been developed in quantum information theory, but most of them are difficult to
compute even for relatively small quantum systems \cite{rmp-review}.
The measure of logarithmic negativity was proposed by Vidal and Werner \cite{vidal01}, as an effective, computable measure of entanglement in a mixed state.  
Partially transposed density matrix can have negative eigenvalues only if the subsystems are quantum entangled.
Although zero negativity does not imply an absence of entanglement in
the sense of separability, it does provide an upper bound on distillable entanglement\cite{vidal01}. Furthermore, its monotonicity properties make it an effective
quantitative measure of how entangled a system is. In pure states it coincides with the Renyi entropy with index $1/2$.

Another motivation for studying log negativity is that we can prove an area-law for it at $T>0$ in arbitrary dimensions, even though we thus far
do not know how to prove an `area-law' for Renyi Mutual Information (MI) for $T>0$. One can regard this 
as a first step toward proving an area law for Renyi MI.

A large body of work exists studying multi-partite negativity in pure systems as well as bi-partite negativity in the thermal state at
nonzero temperatures \cite{plenio,tonni,jvidal,sbose,wang,qmc-negativity}. In models of free particles and harmonic oscillators, the correlation matrix method (aided by Wick's theorem) greatly
simplifies the study and allows for analytical calculations. For spin systems, studies have been restricted mostly to small 1D clusters as exact
calculations of the eigenvalues of the transposed density matrix is a cumbersome task.

In this paper, we show that logarithmic negativity in a lattice model obeys an area-law at all nonzero temperatures.
We also show that it satisfies the linked cluster property. 
However, unlike zero temperature, where negativity is a smooth
function of Hamiltonian parameters except at quantum phase transitions, at nonzero temperatures negativity is not always a smooth function
of temperature. It is characterized by a sudden death, a temperature above which it becomes identically zero. Furthermore, it has multiple
smaller sudden onsets as a function of temperature, which make it unsuitable for any power series expansion \cite{book,series-reviews}. Numerical Linked Cluster (NLC)
methods \cite{nlc,nlc-dmrg,nlc-et} provide an ideal way to calculate logarithmic negativity for a thermodynamic system. We use both exact diagonalization and
NLC to calculate logarithmic negativity for prototypical quantum spin models in one and two dimensions.

We also use perturbation theory to develop a simple physical picture for the onset of negativity 
at a finite temperature (also known as sudden death of negativity when going up in temperature).
Fluctuations in high weight states (such as the ground state) near the interface between subsystems,
upon partial transposition create a large off-diagonal coupling between low weight states.
The mixing between these low weight states results in an entangled Bell pair of states with a negative (and a positive)
eigenvalue for the partially transposed density matrix. Sudden death arises because of the need to build the quantitative difference between large and small
weight states, which goes to zero as $T$ goes to infinity. This non-analytic temperature dependence has nothing to do with any
large length scale in the problem. Even though sudden death temperature is not identical for different clusters, it
converges rapidly enough with system size to lead to a well converged negativity function in the calculations. 

We also find that entanglement negativity is unaffected by classical phase transitions. In some parameter regions, the sudden death happens at a temperature
lower than the phase transition termperature, implying that it remains zero at the phase transition. In other parameter regimes, sudden death happens at a temperature
higher than the phase transition temperature. Development of long-range correlations spoils the convergence of NLC and typically leads to strong
oscillations in the terms, but their sum is still well approximated by Euler summation. The resulting negativity is quite insensitive to the phase transition itself. 
This is consistent
with the idea that quantum mechanics has very little role, at the macroscopic level, in classical phase transitions. This should be contrasted with Mutual Information,
defined in terms of Von Neumann or Renyi entropies, which show a clear signature of the classical phase transitions \cite{melko,series-prl}, and hence, are dominated by classical correlations.

On approach to a $T=0$ quantum phase transition, negativity develops universal critical singularities. At $T=0$, it coincides with the
Renyi entropy with index $1/2$ and in $1+1$-d has the well known logarithmic singularity in the size of the system, with
a coefficient determined by the central charge of the conformal field theory (CFT) \cite{cardy}. At a finite but low temperature, the singularity is
rounded off, in the thermodynamic limit, and instead a saturation to a logarithmic behavior in inverse temperature $\beta$ is observed.

The plan of the paper is as follows. A formal proof of the area-law for log negativity at any nonzero temperature is provided in
the appendix. In the next section we provide the basic definitions and discuss the methods used for the calculations.
That is followed by results for entanglement negativity in one and two dimensional quantum spin models. Finally, the last section provides a discussion
and conclusions.

\section{Models and Methods}

We consider two prototypical quantum spin models. The first is the antiferromagnetic XXZ model, with parameter $\lambda$ defined 
by the Hamiltonian
$${\cal H}= \sum_{\langle i,j \rangle} (S_i^z S_j^z +  \lambda[S_i^x S_j^x + S_i^y S_j^y]).$$
The second is tha transverse field Ising model (TFIM), defined in terms of the parameter $J$
$${\cal H}= -J\sum_{\langle i,j \rangle} S_i^z S_j^z - \sum_iS_i^x. $$ 
We are interested in the entanglement in the thermal state characterized by the density matrix
$$ \rho = \frac{e^{-\beta\cal{H}}}{\cal{Z}}, $$
where ${\cal Z}$ is the partition function.

For a bi-partitioned system, with one partition labeled A, and the other B, the (logarithmic) negativity is defined as
\begin{eqnarray}
%{\cal{N}}&=& \frac{||\rho^{\Gamma_A}||-1}{2} \\
	{\cal{N}}&=& \ln||\rho^{\Gamma_B}||_1, 
\end{eqnarray}
where $\rho^{\Gamma_B}$ denotes the partial transpose of the density matrix
with respect to subsystem B, and $||X||_1$ denotes the trace norm of $X$.  More explicitly, we
write the density matrix acting on the Hilbert space $H_A \otimes H_B$ as
$$\rho = \sum_{ijkl}P_{kl}^{ij}\ket{A_i}\bra{A_j} \otimes \ket{B_k}\bra{B_l}. $$
We introduce an operator $T$ that transposes in the Hilbert space of $\cal{H}_B$, then the partial transposed matrix $\rho^{\Gamma_B}$ is defined as
\begin{eqnarray}
\rho^{\Gamma_B} &=& I \otimes T(\rho) = \sum_{ijkl}P_{kl}^{ij}\ket{A_i}\bra{A_j} \otimes [\ket{B_k}\bra{B_l}]^T\nonumber \\
 &=& \sum_{ijkl}P_{kl}^{ij}\ket{A_i}\bra{A_j} \otimes \ket{B_l}\bra{B_k} \nonumber
\end{eqnarray}
with $I$ being the Identity operator in the Hilbert space $\cal{H}_A$. Note
that $\rho^{\Gamma_A} =\rho^{\Gamma_B}$, for a real symmetric density matrix.

In a pure state, one can use the Schmidt decomposition of a wavefunction to show that \cite{vidal01} 
\begin{eqnarray}
{\label{negschmidt} 
||\rho^{\Gamma_B}||_1 = \left[\sum_{\alpha}{C_{\alpha}}\right]^2
}
\end{eqnarray}
where $\{C_{\alpha}\}$ are the schmidt coefficients.  
%If we consider the reduced
%density matrix $\rho_A$ for a pure state $\ket{\Psi}$, obtained by tracing out B degrees of freedom from the full density matrix
%$$\rho_A = \substack{\Tr\\ B}(\rho) = \substack{\Tr\\ B}(\ket{\Psi}\bra{\Psi})$$
%if working in the Schmidt basis it is simple to show
%\begin{eqnarray}
%{\label{reddense} 
%\rho_A = \sum_{\alpha}C_{\alpha}^2\ket{A_{\alpha}}\bra{A_{\alpha}}
%}
%\end{eqnarray}
%implying that the eigenvalues of $\rho_A$ are the squares of the schmidt coefficients. So to compute negativity for a pure state, we constructed and numerically diagonalized $\rho_A$ and utilized equations (1, \ref{negschmidt} \& \ref{reddense}).
Another set of measures of entanglement based on the reduced density matrix $\rho_A$ are the well studied Renyi entropies of order $\alpha$ defined as
$$ S_{\alpha} = \frac{1}{1-\alpha}\ln\left[\Tr\left(\rho_A^{\alpha}\right)\right]$$
It is easy to see that negativity in the pure state is equal to Renyi entropy of index $1/2$.

\subsection{Linked cluster method}
The linked cluster method\cite{nlc-et} is based on expanding a property for a larger cluster $\cal{O}$ in terms of all its subclusters c.
$$ \cal{P}(\cal{O}) = \sum_{\text{c} \subseteq \mathcal{O}} W(\text{c}).$$ 
The weights of the subclusters are defined recursively by the subgraph subtraction procedure:
$$ \cal{W}(\cal{O}) = P(\cal{O}) - \sum_{\text{c} \subset \mathcal{O}} W(\text{c}).$$
In a thermodynamic system, with translational symmetry, one can combine all clusters with translational symmetry.
Provided a quantity satisfies the linked-cluster property, {\it i.e.} only linked clusters have non-zero weight,
it can be calculated by simple graphical methods and leads to an appropriate extensive or intensive quantity in the
thermodynamic limit.

For the logarithmic negativity, we can show that only connected clusters shared between subsystems A and B have non-zero weight.
It is evident that unless a cluster is shared between A and B it can not have any negativity. To show that a disconnected cluster has
zero weight, we first observe that for a disconnected cluster  $c_1 U c_2$ both of which ($c_1$ and $c_2$) may have parts in A and B, the partially transposed
density matrix can be written as a product of partially transposed density matrices over the two clusters. Thus, the eigenvalues of the partially
transposed density matrix would be a product of eigenvalues for the two clusters. It follows that logarithmic negativity will be a sum
of the logarithmic negativity over the two clusters and upon subgraph subtraction, it will have zero weight. If we imagine
partitioning a large system into two halves by a planar partition, the clusters will have translational symmetry perpendicular to
the partition.  The count of each cluster modulo such a translation
will be proportional to the area of the interface. As a result, the lograithmic negativity per unit area can be expressed as
$$ N= \sum_{\text{c}} W_N(\text{c}),$$ 
where $W_N(\text{c})$ is the weight of the cluster c and the sum is over all translationally distinct clusters of the lattice. 

The NLC calculation requires evaluation of negativity for clusters with free boundary conditions (clusters that actually occur in the lattice).
In one-d, we also do exact diagonalization for clusters with periodic boundary conditions as that provides an alternative method for studying the
large size limit. On desktop computers, we can evaluate the finite-temperature negativity for up to 14-site clusters, using a brute-force diagonalization
method. However, since we need to do a large number of such calculations, most of our results are based on systems of size 12 or smaller.
For the ground state properties, we have evaluated negativity for
up to $26$ site clusters using the Lanczos method. The two calculations must agree in the $T$ going to zero limit for each cluster,
provided the cluster has a non-degenerate ground state. 
This comparison provides a non-trivial check on our numerical calculations.
Note that in XXZ systems with odd number of spins, we have  Kramers degeneracy, and the
two calculations give different results, as at any nonzero temperature the density matrix is always mixed.

\begin{figure}[h!]
\begin{center}
\includegraphics[width=\linewidth]{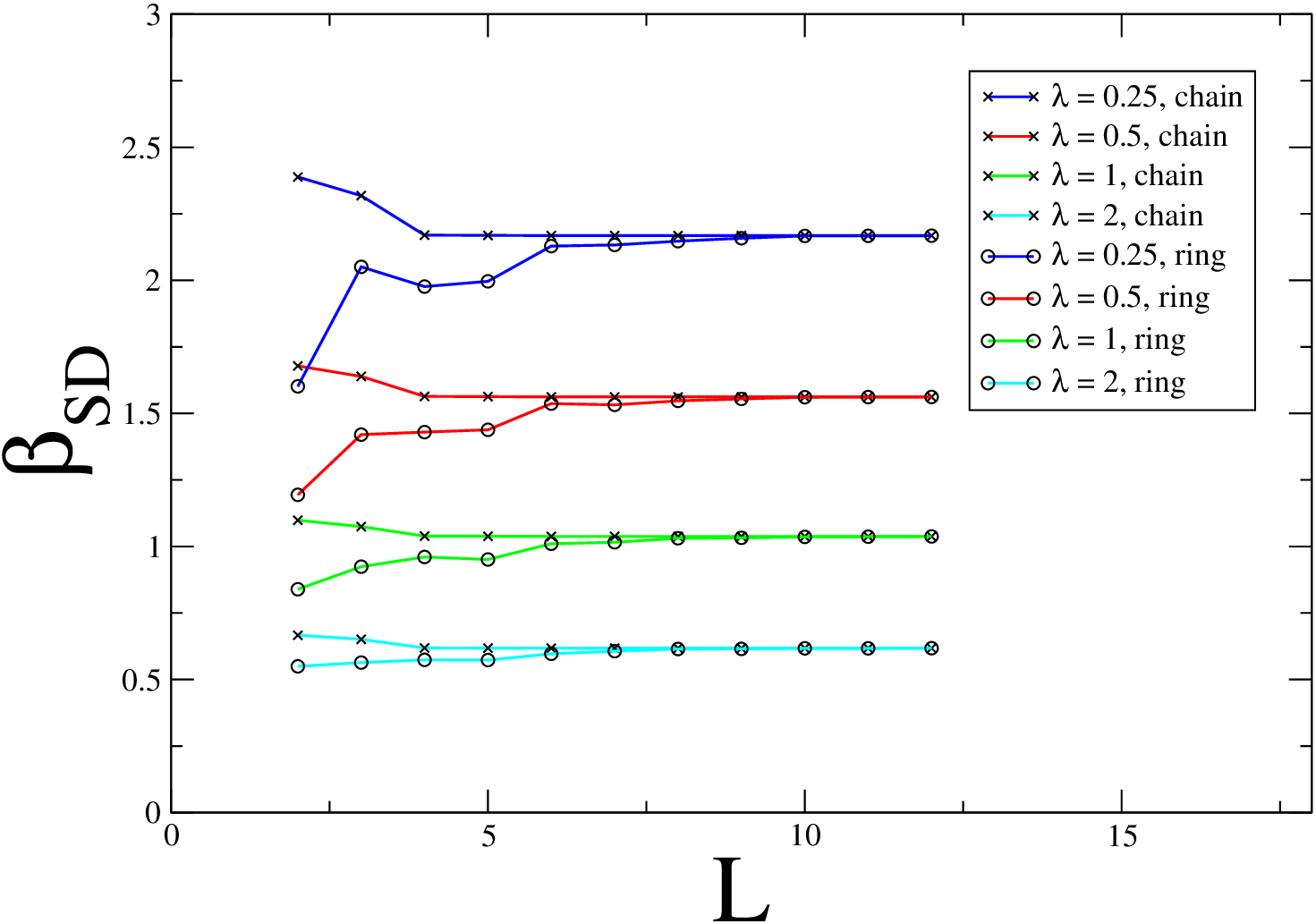} 
\caption{Sudden death inverse temperate $\beta$ values for the antiferromagnetic XXZ models for different size rings and chains.
Note that the results converge from above for the chains and from below for the rings.
}
\end{center}
\end{figure}

\begin{figure}
\begin{center}
\includegraphics[width=\linewidth]{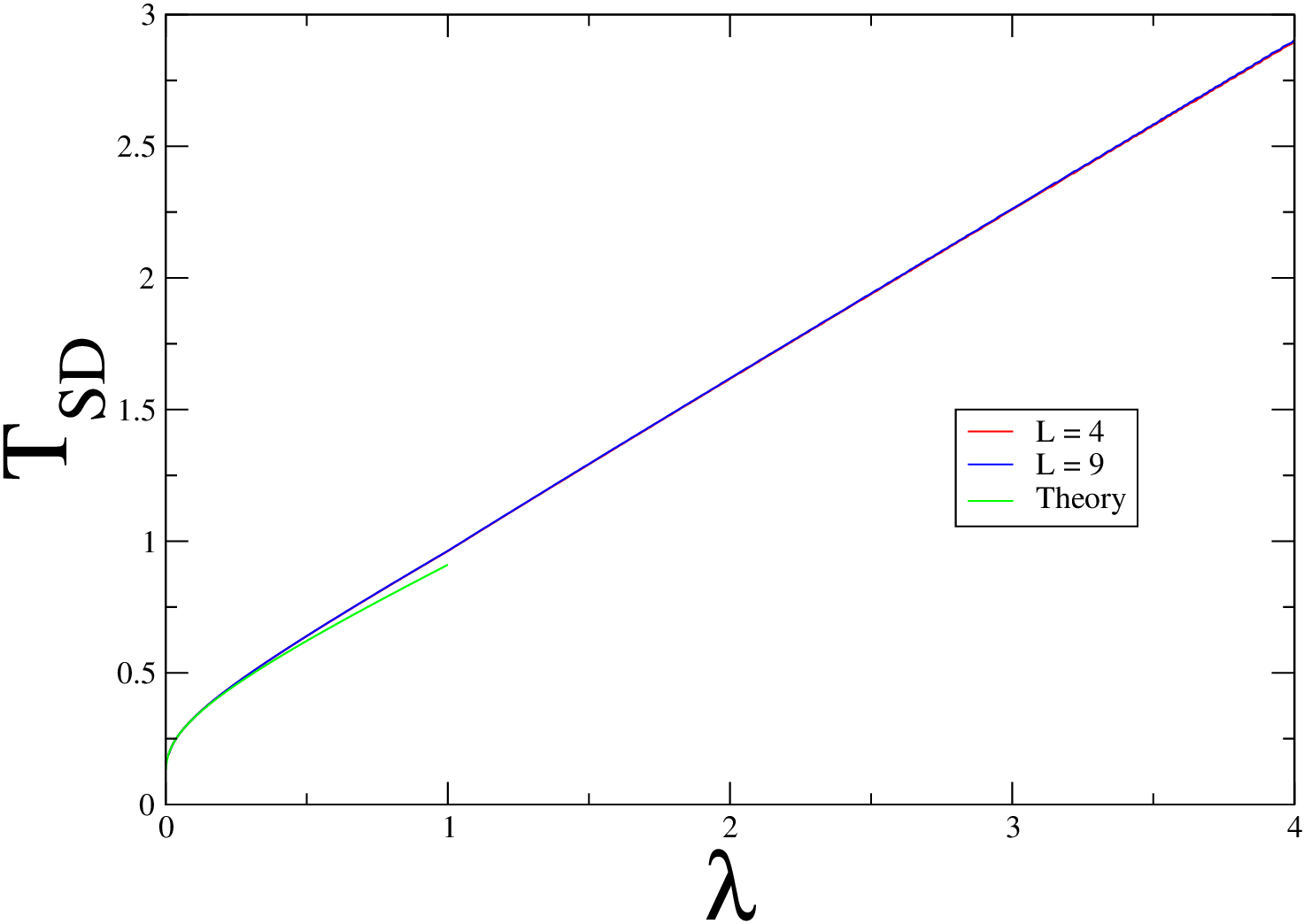} 
\caption{\label{sdt} Comparison of sudden death temperature $T_{SD}$ for chains of length $4$ and $9$ with the perturbation theory results.
}
\end{center}
\end{figure}

We begin with one dimensional models.
Our interest is in calculating the logarithmic negativity $N$ when a large chain is partitioned at a bond.
As the area between subsystems is just a point in 1d, an area-law means finite entanglement in the thermodynamic limit.
We can do this calculation in two ways. We can consider a chain of length L with periodic boundary conditions and divide it into two equal parts A and B 
and calculate the logarithmic negativity for this partition. If the entanglement arises only from the area,
in the limit of $L$ going to infinity, this should equal twice the entanglement of each cut. Alternatively, we can use the NLC method. In this case,
in each order ($n$) there is only one topological cluster at each order, an open chain of length $n$. However, we need to place the partition on every bond of the 
cluster and sum them up. This
amounts to positioning the cluster at different places with respect to the cut. The NLC method will sum to the negativity $N$.

\begin{figure}
\begin{center}
\includegraphics[width=0.5\textwidth]{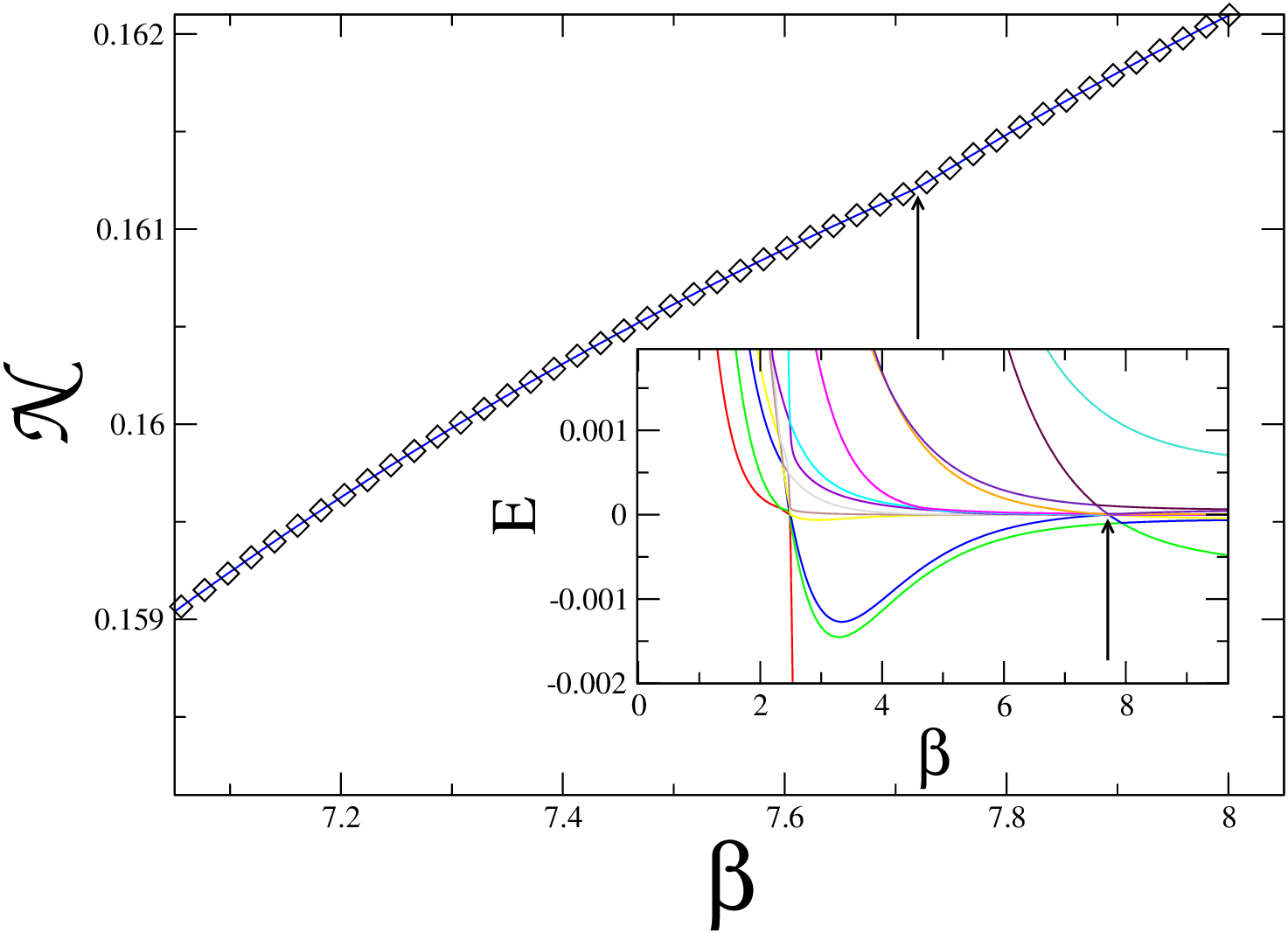} 
\caption{\label{second-onset} A second onset in negativity happens for a $4$-site system below an inverse temperature of $\beta=8$. 
The inset shows eigenvalues of the partially transposed density matrix as a function of $\beta$. Note that in a small window near these
onsets there are multiple eigenvalues which change sign (going from positive to negative and from negative to positive).
}
\end{center}
\end{figure}

\section{Results}
\subsection{Sudden death}

It is well known that negativity in the thermal state becomes zero above a sudden death temperature $T_{SD}$.
We would like to understand this sudden death phenomenon in our lattice models.
Fig.~1 shows the sudden death $\beta$ values for different strengths of the coupling $\lambda$ in the XXZ model for rings of length $L$. It is apparent that
this converges rapidly with $L$, implying that it is a short-distance property. 
This motivates that perhaps sudden death may be explained well by low order perturbation theory.
In fact, this sudden death can already be seen in first order perturbation theory.

Let us look at the 1D XXZ model with small $\lambda$.
We treat $\lambda$ in the XXZ Hamiltonian as a small parameter, and expand around $\lambda=0$.
Negative eigenvalues in $\rho^{\Gamma_B}$ arise from elements in the off-diagonal having a larger magnitude than its corresponding diagonal elements.
The large off-diagonal elements that cause negativity in the XXZ model comes as a result of states with high Boltzmann weight, that is typically the ground state.
At $\lambda=0$, there are two degenerate ground states which are simply product states of spins ordered antiferromagnetically along $z$, which do not mix in the thermodynamic limit.

The perturbation acting on these states flips two spins, creating a pair of excited bonds with total energy $E_0+1$, where $E_0$ is the unperturbed ground state energy.
Noting that in the partial transposition of $\rho$, off-diagonal elements which consist of only a pair of spins being flipped far away from the boundary are not affected, which means we must only consider the two spins right on the boundary being flipped.
This means that, defining $\rho_0$ to be the pure state density matrix of the ground state, the only relevant off-diagonal element in the partial transpose (where $\colon$ represents the boundary between $A$ and $B$, and only the 4 spins nearest to the boundary are included) is
\begin{equation}
    \left<\uparrow \downarrow : \uparrow \downarrow \right|\rho_0 \left| \uparrow \uparrow : \downarrow \downarrow \right>
    =
    \left<\uparrow \downarrow : \downarrow \downarrow \right|\rho_0^{\Gamma_B} \left| \uparrow \uparrow : \uparrow \downarrow \right>
    = -\lambda/2
    \label{}
\end{equation}
and its transposed elements.

As we are interested in the sudden death $\beta_\text{SD}$, which should be very large when $\lambda$ is small, we can assume that other offdiagonal elements are negligible compared to those from the ground states.
In the partial transpose of the thermal density matrix $\rho(\beta)=\exp(-\beta \mathcal{H})/\mathcal{Z}$, the offdiagonal element becomes
\begin{equation}
    \left<\uparrow \downarrow : \downarrow \downarrow \right|\rho^{\Gamma_B} \left| \uparrow \uparrow : \uparrow \downarrow \right>
    =
    -(\lambda/2) (e^{-\beta E_0} - e^{-\beta [E_0+1]})/\mathcal{Z}
    \label{}
\end{equation}
while the diagonal elements are
\begin{equation}
    \left<\uparrow \downarrow : \downarrow \downarrow \right|\rho^{\Gamma_B} \left| \uparrow \downarrow : \downarrow \downarrow \right>
    =  
    \left<\uparrow \uparrow : \uparrow \downarrow \right|\rho^{\Gamma_B} \left| \uparrow \uparrow : \uparrow \downarrow \right>
    =
    e^{-\beta [E_0+1]}/\mathcal{Z}
    \label{}
\end{equation}

Thus, the sudden death $\beta_\text{SD}$ where the eigenvalue in $\rho^{\Gamma_B}$ becomes negative, obtained by solving the 2 by 2 matrix, is simply given by
\begin{align}
    \beta_\text{SD} =  \ln(\frac{\lambda + 2}{\lambda})
\end{align}
We see that the sudden death temperature goes to 0 for $\lambda=0$, and actually does so logarithmically $T_\text{SD}\approx 1/\ln(2/\lambda)$.
This is true also for the TFIM in the small $J$ limit, which can be shown in a similar manner in leading order perturbation theory.

Fig.~2 shows the calculated sudden death temperature as a function of $\lambda$, in very good agreement with first order perturbation theory at low $\lambda$. 
As expected the sudden death temperature goes to zero as $\lambda$ goes to zero logarithmically. 
We do not show $T_{SD}$ for the transverse field Ising model
here, as we will discuss its sudden death in context of finite temperature phase transitions in
two dimensional models. The sudden death phenomena is qualitatively similar for our models
and does not depend much on dimensionality. 

It is interesting to note that, in general, models can have multiple sudden discontinuities in the derivative of their log negativity as more and more
eigenvalues of the partially transposed density matrix turn negative. 
For example, the negativity for a chain of length $4$ cut in two halves has a discontinuity in the slope at $\beta$ just below $8$ as shown in Fig.~3, in additional to the first discontinuity at $\beta_\text{SD}$.  
At this point, an eigenvalue in $\rho^{\Gamma_B}$ (which arise in 3rd order perturbation theory) becomes negative.
These discontinuities persist into the thermodynamic limit, but contributions from higher order ones are usually much smaller compared to the first one.

\subsection{Area laws}

\begin{figure}
\begin{center}
\includegraphics[width=0.5\textwidth]{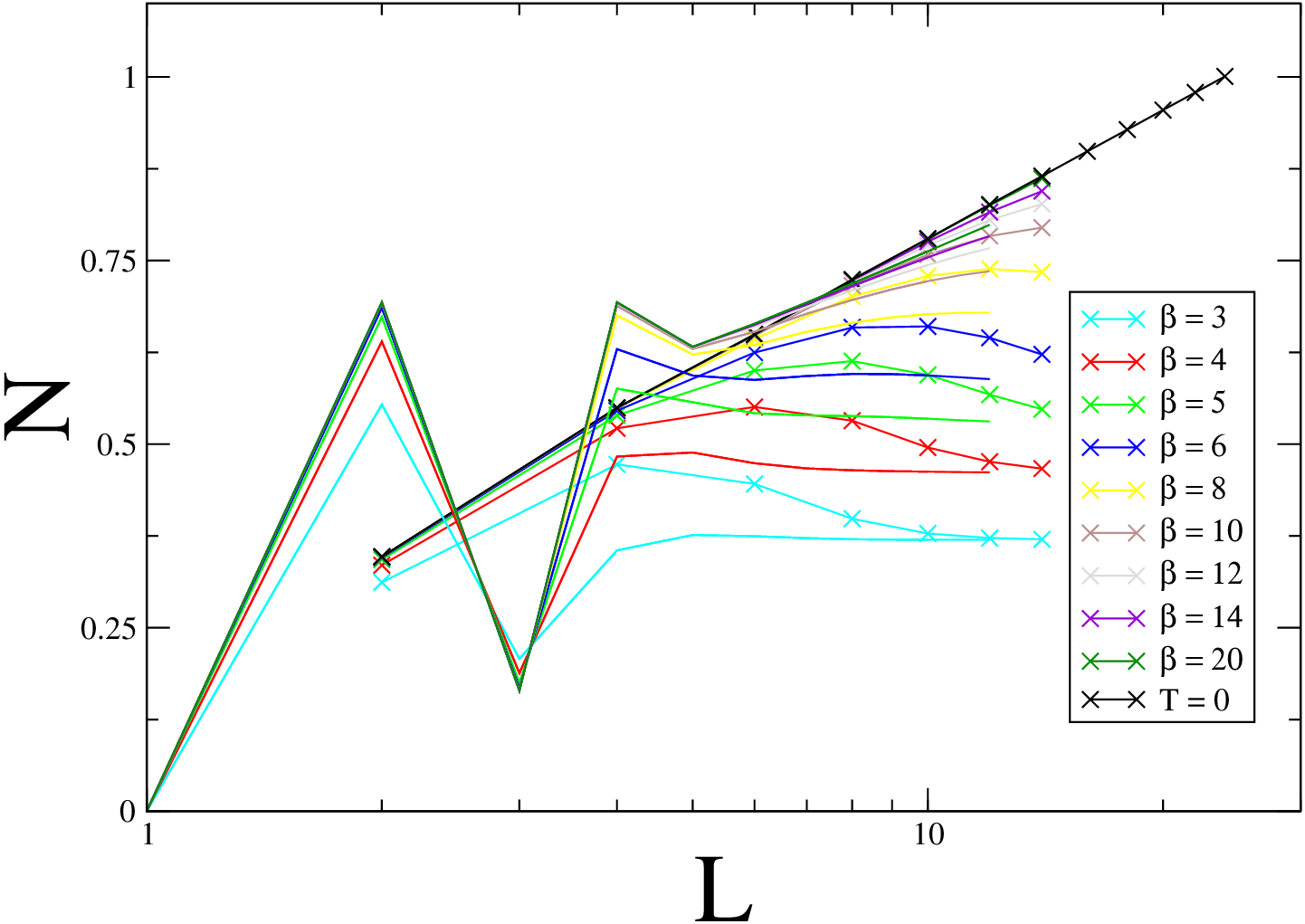} 
\caption{\label{scaling1} 
Logarithmic negativity for the antiferromagnetic Heisenberg model ($\lambda=1$)  as a function of the length $L$ from
exact diagonalization (ED) of periodic chains 
at different $\beta$ values, as well as from NLC up to order $L$. 
The terms in the NLC series are strongly alternating, so an Euler summation is used starting from the 4th term to smoothen the sum.
Also shown are the results from the Lanczos based calculation for the ground state negativity.
Plots with symbols are from ED while those without symbols are from NLC.
}
\end{center}
\end{figure}

\begin{figure}
\begin{center}
\includegraphics[width=0.5\textwidth]{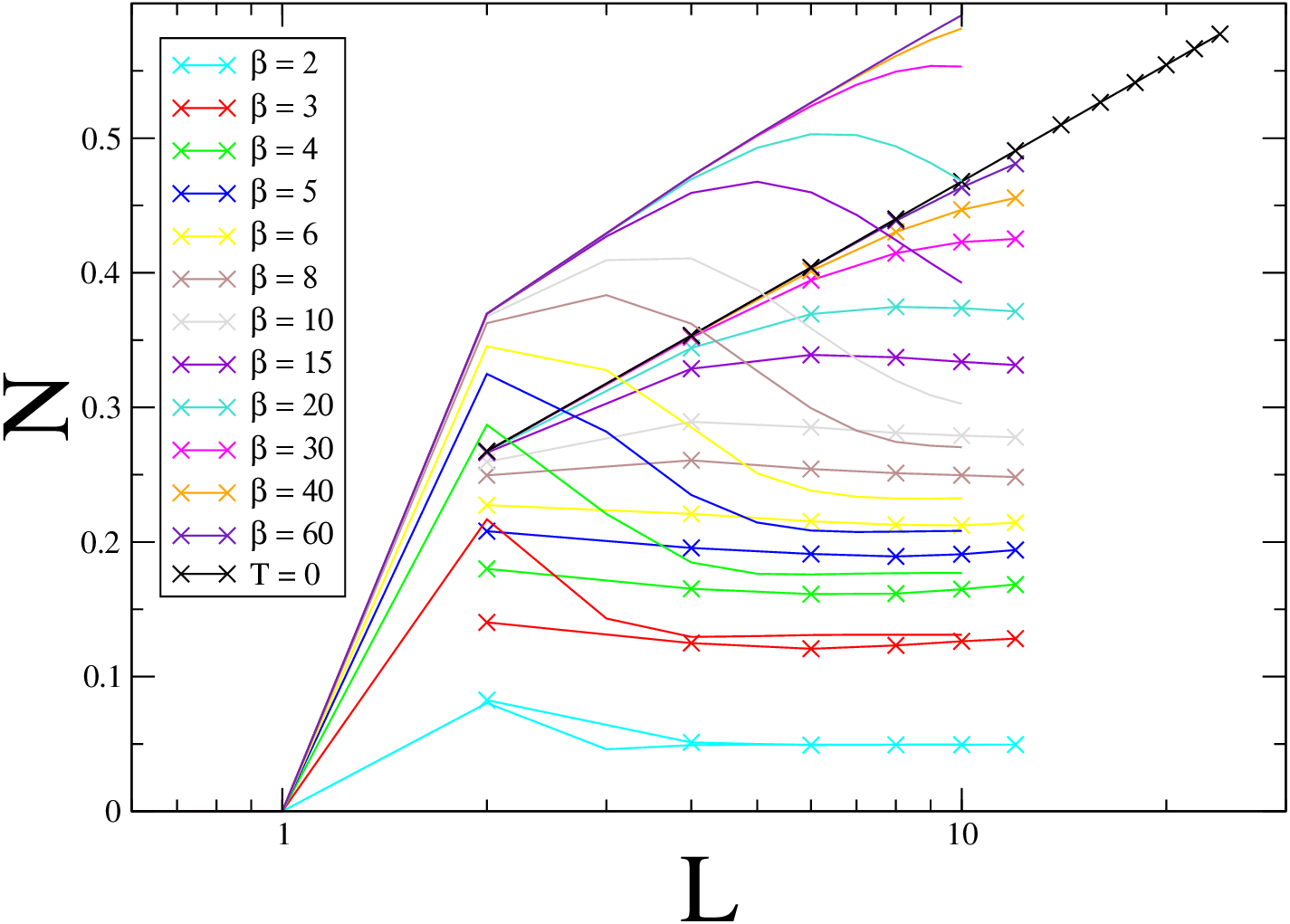} 
\caption{\label{scaling2} 
Logarithmic negativity for the transverse-field Ising model (TFIM) tuned to its quantum critical point ($J=2$)  as a function of the length $L$ from
exact diagonalization (ED) of periodic chains and also NLC up to order $L$ for different $\beta$ values. Also shown are the results from the Lanczos based calculation for the ground state negativity.
Plots with symbols are from ED while those without symbols are from NLC.
}
\end{center}
\end{figure}

\begin{figure}
\begin{center}
\includegraphics[width=0.5\textwidth]{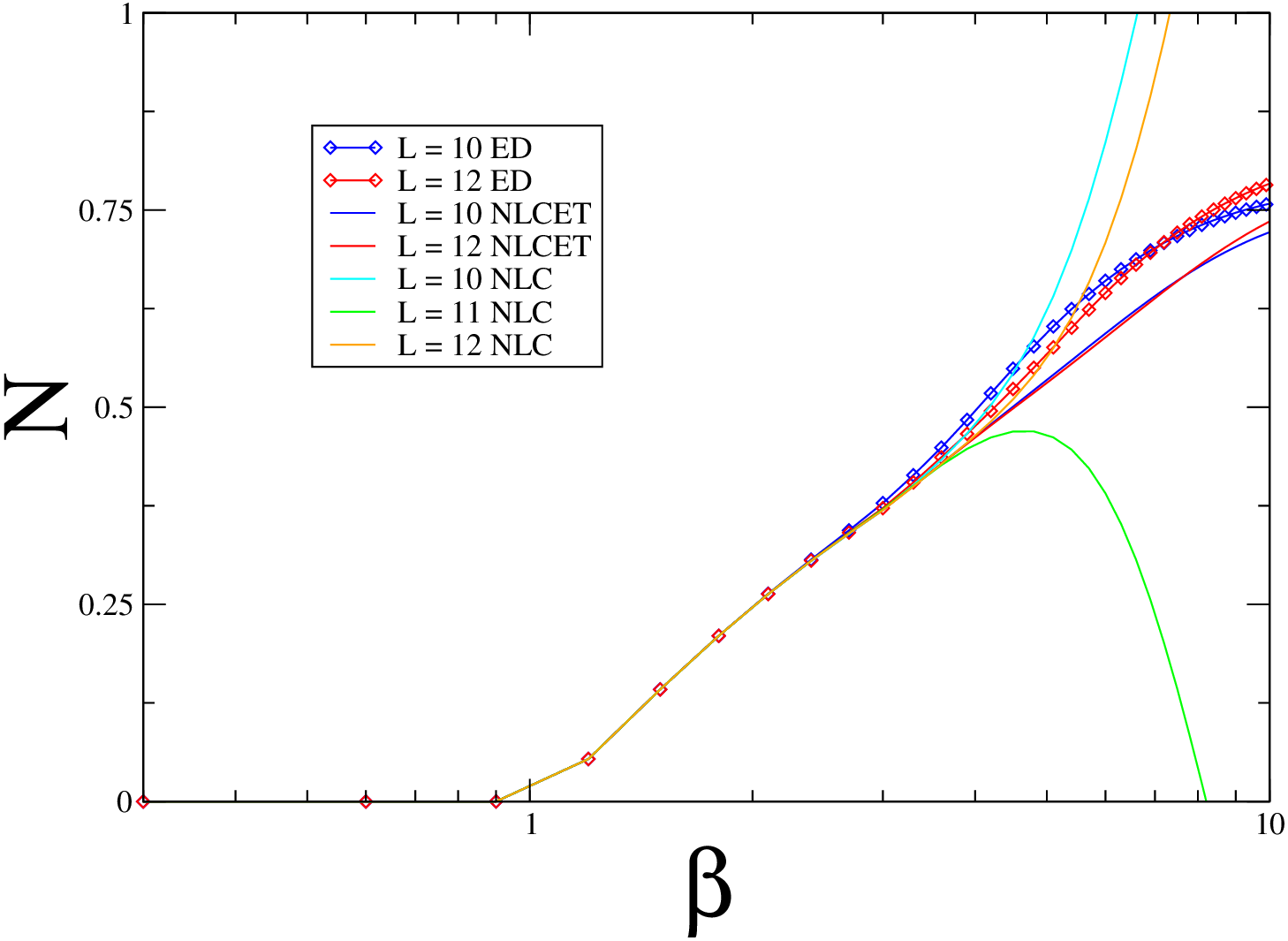} 
\caption{\label{nvbpbc} 
Logarithmic negativity of the antiferromagnetic Heisenberg model as a function of $\beta$ calculated by exact diagonalization
of rings of size $L$ and by NLC. For the NLC the partial sums of different orders and the Euler sums are shown.
}
\end{center}
\end{figure}

\begin{figure}
\begin{center}
\includegraphics[width=0.5\textwidth]{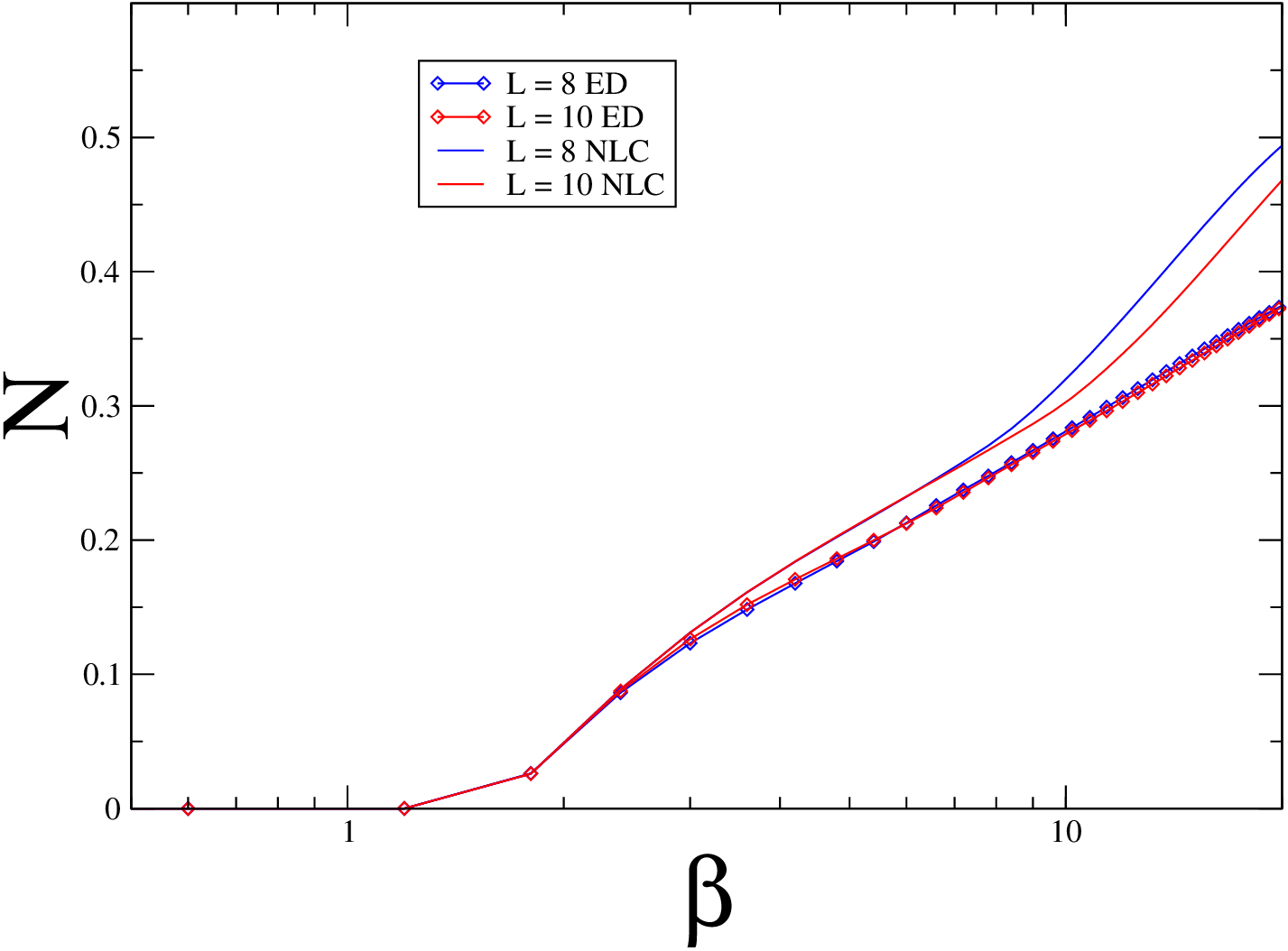} 
\caption{\label{nvbj2} 
Logarithmic negativity for the transverse field Ising model (TFIM) tuned to its quantum critical point ($J=2$)  as a function of $\beta$ calculated by exact diagonalization
of rings of size $L$ and by NLC. For the NLC the partial sums of different orders are shown.
At intermediate $\beta$, ED has not yet converged and appears to be increasing towards the NLC value.
}
\end{center}
\end{figure}

It is well known that entanglement in the ground state of gapped Hamiltonians follow an area law \cite{wolf08}.  
In 1d critical or gapless systems, the ground state entanglement entropy instead diverges logarithmically with size of the system. 
A natural question concerns the fate of negativity at non-zero temperatures in this situation, does it follow an area law despite the logarithmic entanglement in the ground state? As shown in the appendix, logarithmic negativity always satisfies an area-law at nonzero temperatures.

At a more quantitative level, we study negativity as a function of the length of the periodic chains $L$ at different inverse temperatures.
Fig.~4 and Fig.~5 shows negativity as a function of the length of the periodic chains $L$ at different inverse temperatures
for the antiferromagnetic Heisenberg model (with $\lambda=1$) and the TFIM at its critical point ($J=2$). 
Quantum entanglement in these models is largest at $T=0$ and is usually a monotonically decreasing function of temperature \cite{footnote1}.
At $T=0$ the logarithmic behavior of the negativity is clearly apparent at our sizes shown in the line labeled $T=0$. 
For higher temperatures, on the other hand, the negativity appears to increase initially but eventually saturate to a constant value.
Also shown are the results from NLC, which for low $\beta$ are converging faster than ED.
These results confirm that negativity indeed follows an area law at nonzero temperatures, even when the ground state negativity does not.

In Fig.~6 and Fig.~7 we show the results for the logarithmic negativity as a function of inverse temperature $\beta$ for the Heisenberg model
and the TFIM at their critical point. 
Results from exact diagonalization as well as from NLC are shown. 
It is clear that the results
converge rapidly at high temperatures between the exact diagonalization and NLC. As the temperature is lowered the system develops a logarithmic
behavior in $\beta$ as expected from scaling theory. 
At low temperatures, as the quantum critical point is approached, the convergence must break down as
larger system sizes would be needed to obtain the results in the thermodynamic limit. 
We find that the convergence is faster for the Heisenberg model than for the TFIM,
until the terms start to oscillate strongly with order.
In this case, the Euler summation greatly improves the  convergence of NLC.
For intermediate $\beta\approx5$ in Fig~\ref{nvbj2}, NLC has converged while the ED estimates are still increasing with order, as can be seen from Fig~\ref{scaling2}.
However, the logarithmic  behavior of negativity in $\beta$ is apparent in both cases.

We can use a simple scaling argument to relate the coefficient of the logarithm in $L$ at $T=0$ and the coefficient of logarithm 
in $\beta$ at finite temperatures in the thermodynamic limit
to each other. In the ground state of one-dimensional critical systems, for arbitrary Renyi index $\alpha$, the logarithmic singularity for a single cut, for
a model with central charge $c$, is known to take the form\cite{cardy}
\begin{equation}
N={c\over 12} (1+{1\over \alpha}) \ln{L} + constant.
\end{equation}
For $\alpha=1/2$, appropriate for the the log negativity, the coefficient for the logarithm becomes $c/4$.

Our fits for the TFIM and the Heisenberg model at $T=0$ give coefficients of  $\ln{L}$ of $0.125$ and $0.25$ respectively, in excellent agreement with the
known central charge values of $c=1/2$ and $c=1$. Finite $\beta$ implies that the imaginary time direction has extent $\beta$.
Thus, by Lorentz symmetry of the critical point, $\beta$ should play the role of the size of the system
and one would expect the same coefficient for $\ln{\beta}$ as well. In our fits we find the coefficients of $\ln{\beta}$ to be $0.13$ and $0.32$ for the TFIM
and Heisenberg models respectively. Given that our results are not converged at very low temperatures, and linearity in $\ln{\beta}$ may be slowly
changing with $\beta$, these results are consistent with the scaling argument.

\subsection{Bipartite logarithmic negativity in two dimensions}

We now turn to the two-dimensional square-lattice. We are interested in the logarithmic negativity when a large system is divided into two halves
by bisecting a set of parallel bonds. In this case, an area-law obeying entanglement negativity will be proportional to the length of the line, and one
can calculate the entanglement negativity per unit length.
In this case the exact diagonalization method is no longer so useful, but we can continue
to use the numerical linked cluster expansions to get to the results in the thermodynamic limit. We will focus on the transverse-field Ising model, which has a larger Hilbert
space due to lack of conservation laws, but also provides a model with both an ordered and a disordered phase. 

The ground state of the square lattice TFIM model has a quantum critical point at $J\approx0.657$, where entanglement entropy and negativity exhibit a singularity.  
Standard NLC expansions are unable to accurately see this singularity, however, as convergence is very poor in the ordered phase.  
This is due to the Schrodinger cat states which contribute an additional entanglement of $\ln(2)$ independent of the cut area, and makes convergence difficult in our NLC calculation of entanglement per unit area.
To get around this, we employ a modified NLC where each cluster is assumed to be embedded in a system of ferromagnetically frozen spins (also called a ``low temperature'' expansion).
This effectively acts as an applied field on the boundary of the cluster, destroying the cat states.

We are interested in how  the entanglement persists to non-zero temperatures, when the phase transition is classical in nature.
Quantum entanglement should play no role, and hence we would also expect that negativity should show no signs at the classical critical point.
Fig.~8 shows negativity as a function of $J$ for various temperatures up to orders 7 and 8, as well as the expected critical couplings $J_c$.
At $T=0$, NLC clearly shows a maximum that is getting sharper near the critical point.  
Because this expansion favors the ordered phase, the peak occurs slightly below $J_c$, but is getting closer with order.
At higher temperatures, negativity is zero below a certain $J$ because of the sudden death phenomena, and so the peak is shifted to higher $J$.  
The expansion appears essentially converged for these curves, and the location of the peak is uncorrelated with the classical transition points, confirming our expectation that negativity plays no role in classical phase transitions.
In fact, beyond a certain point, negativity is zero at the classical phase transition.
Fig.~\ref{figure11} shows the sudden death temperature for the model, which intersects with the classical phase transition point.

\begin{figure}
    \includegraphics[width=0.5\textwidth]{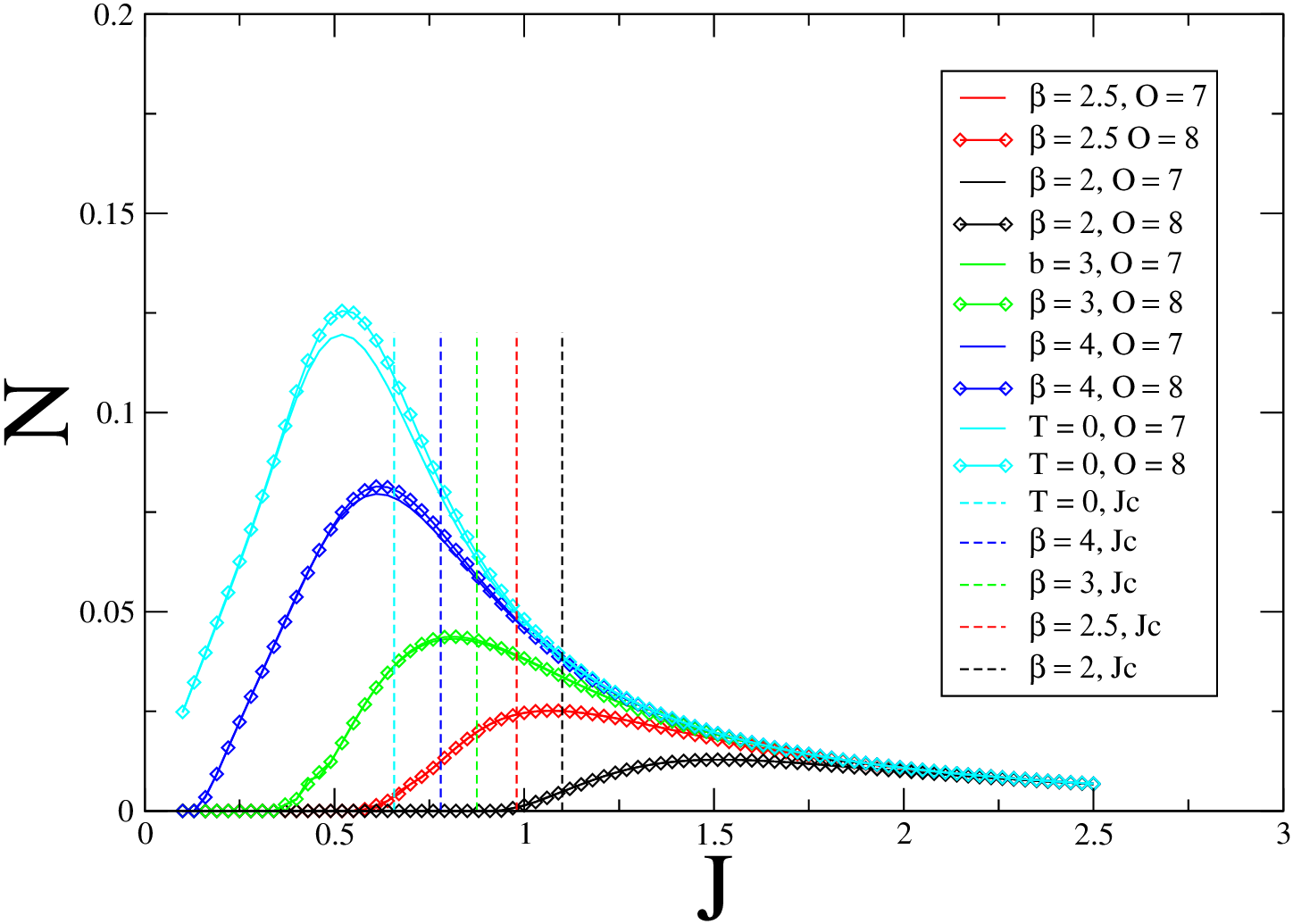}
    \label{nvjlt}
    \caption{Logarithmic negativity for the TFIM on a square lattice as a function of $J$ for varying temperatures from the
``low temperature'' expansions (see text).}
\end{figure}
%\begin{figure}
%\centering
%\includegraphics[width=0.49\linewidth]{plotj01.eps} 
%\includegraphics[width=0.49\linewidth]{plotj05.eps}

%\includegraphics[width=0.49\linewidth]{plotjc.eps}
%\includegraphics[width=0.49\linewidth]{plotj08.eps}
%\caption{\label{figure10} Logarithmic negativity for the transverse-field Ising model (TFIM) on a square lattice as a function of inverse temperature $\beta$, obtained via NLC. Multiple values of $J$ are shown with (a) $J=0.1$, (b) $J=0.5$, (c) $J=0.657 (J_c)$, and (d) $J=0.8$.  
%}
%\end{figure}

%Fig.~10 shows a plot of the entanglement negativity per unit length for TFIM at various values of $J$. Fig. 11 shows the phase diagram for the two-dimensional
%model. For $J<J_c\approx 0.657$, the system stays disordered down to $T=0$. At larger $J$ values there is a ferromagnetically ordered phase at low temperatures.
%We can see that the NLC converges very well at small J. As J is increased towards the critical point, the convergence is very good up to a second onset point,
%after which higher order terms become strongly oscillatory. We use Euler summation to sum such a series and this appears to converge reasonably well.

\begin{figure}
\centering
\includegraphics[width=0.5\textwidth]{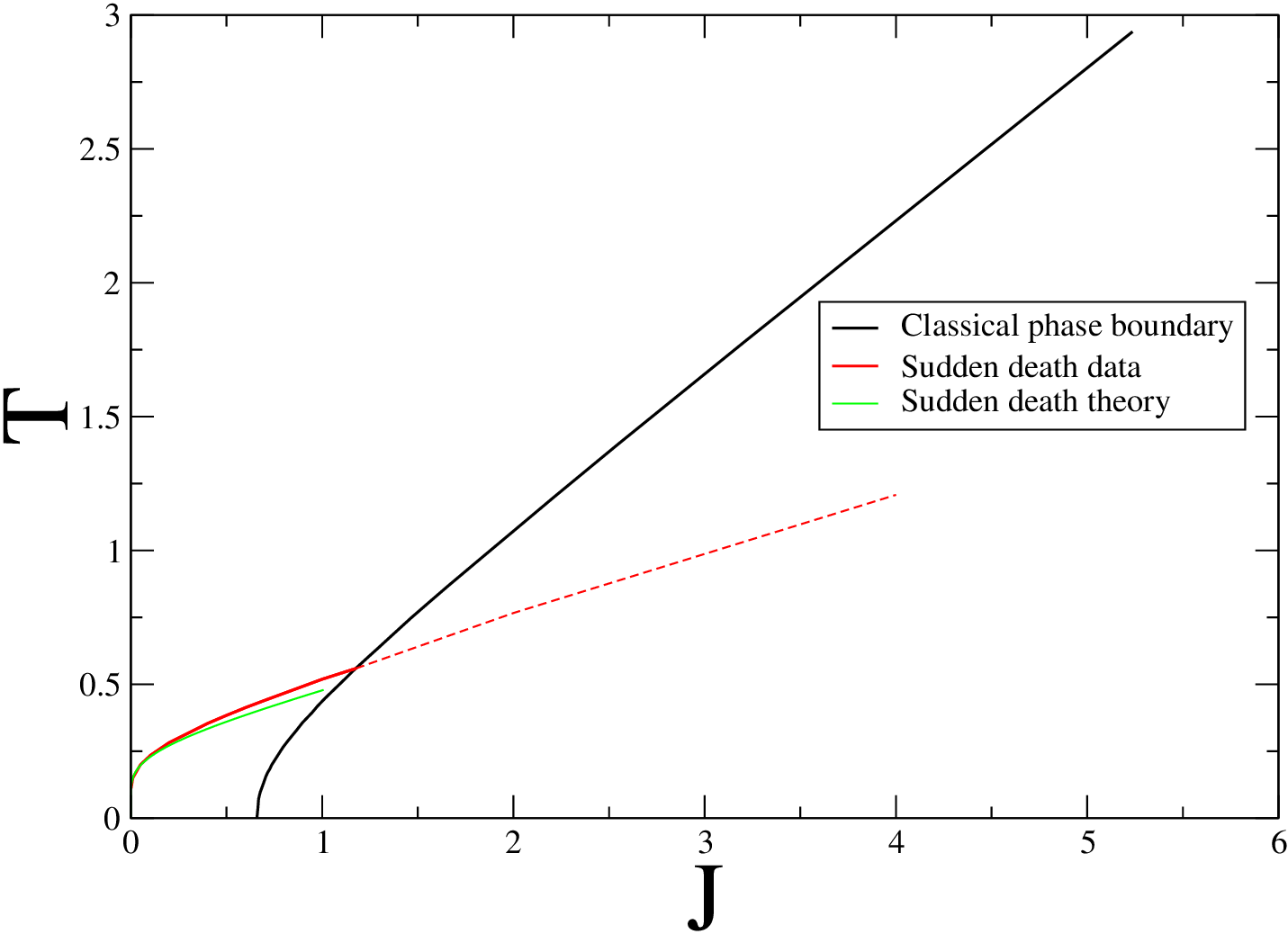} 
\caption{\label{figure11} Phase diagram for the transverse-field Ising model (TFIM) on the square lattice.  Also illustrated are the sudden death temperatures found via NLC. Also shown is first order perturbation theory results. Data for the classical phase boundary gathered from \cite{2dising}.
}
\end{figure}

\section{Discussions and Conclusions}

In conclusion, in this paper we have studied logarithmic negativity for quantum spin models using exact diagonalization in one dimension and
the Numerical Linked Cluster method in both one and two dimensions. The models studied were the antiferromagnetic XXZ model and the transverse field Ising model. We
showed the existence of an `area-law' for the logarithmic negativity and also showed that it obeys the linked cluster property.  This allows us to use NLC to calculate this quantity. We found that for these models,
the logarithmic negativity decreases monotonically with temperature. This is in contrast to Von Neumann entropy in an eigenstate of the system which shows a Volume law and increases as one goes to states relevant to higher temperatures.
A relatively small value of quantum entanglement in the highly mixed thermal state is perhaps also an example of monogamy of entanglement. One can regard the thermal
state as a pure state of a larger system with a thermofield double \cite{feiguin}, where one has introduced twice as many degrees of freedom as in the original system.
The strong entanglement between the system and the additional ancilla degrees of freedom at nonzero temperatures
strongly limits its entanglement within different parts of the system making it essentially local, hence the area-law.

Perturbation theory was used to elucidate the onset of entangement negativity at a sudden death temperature $T_{SD}$.
It also showed that Entanglement negativity is an interfacial phenomena that arises from local fluctuation in the ground and low lying states
across the interface between subsystems, which under partial transposition leads to fully entangled pair of states.

We found that quantum entanglement does not play a role in classical phase transitions. Logarithmic negativity can be zero at the transition.
Or, it can be non-zero with no clear evidence for a singularity in it at the transition. At a $T=0$ quantum phase transitions, it develops universal critical behavior in 
temperature and size of the system. The well known log singularity of CFT is rounded off at any nonzero temperature, leaving a logarithmic
dependence on inverse temperature $\beta$. In future, it may be interesting to study multipartite entanglement negativity using such computational methods.

\begin{acknowledgements}
This work is supported in part by the National Science Foundation grant number DMR-1306048 and
by its Research Experience for Undergraduates (REU) program under grant number NSF PHY-1263201.
\end{acknowledgements}

\begin{widetext}

\section{Appendix: Proof of Area Law}

We consider a system made of two subsystems, called $A,B$.
We consider a Hamiltonian $H=H_A+H_B+H_{AB}$, where $H_A$ and $H_B$ are supported on $A,B$ respectively.  Our goal is to bound the log negativity of
the density matrix of this system at any nonzero temperature.
Let
\be
\rho=\cZ^{-1} \exp(-\beta H),
\ee
where $\cZ$ is a normalization constant chosen so that ${\rm tr}(\rho)=1$.
So, we wish to bound
$$\log(|\rho^{\Gamma_A}|_1),$$
where the superscript $\Gamma_A$ denotes a partial transpose on subsystem $A$ and where $| \ldots |_1$ denotes the trace norm (this is equal to the sum of singular values of an operator; for a Hermitian operator this is the sum of absolute values of eigenvalues).

Any term $H_{AB}$ can be written as a sum of products,
\be
\label{sumprod}
H_{AB}=\sum_\alpha H_A^\alpha H_B^\alpha,
\ee
where $H_A^\alpha,H_B^\alpha$ are supported on $A,B$ respectively.  Suppose that this is possible with at most $K$ total terms in the sum,
with $\Vert H_A^\alpha \Vert \Vert H_B^\alpha \Vert \leq J$, for some constant $J$.
Then, our main result in this appendix is the result that
\be
\label{bnd}
\log(\rho^{\Gamma_A}) \leq \beta*(JK+\Vert H_{AB} \Vert).
\ee
where $\Vert \ldots \Vert$ denotes the operator norm (the largest singular value of an operator).  Note that $\Vert H_{AB} \Vert \leq JK$ by a triangle inequality.

To give an application of this result, consider a many-body system with a Hamiltonian that is a sum of local terms so that each term acts on some set with a bounded number of sites with a bounded Hilbert space dimension on each site.
Then, each term which is supported on both $A$ and $B$ can be written as a sum of products as in Eq.~(\ref{sumprod}) with a bounded number of terms in the sum (if the Hilbert space dimension on each site is bounded by $d$ and a term acts on $n_A$ sites in $A$ and $n_B$ sites in $B$, then the number of terms in the sum is as most $d^{2\cdot {\rm min}(n_A,n_B)}$).
If the number of terms supported on both $A$ and $B$ is proportional to the boundary area of $A$, then summing over terms, we can write $H_{AB}$
as in Eq.~(\ref{sumprod}) with the number of terms $K$ proportional to the boundary area of $A$.
If each term in the Hamiltonian is bounded, we can bound the operator norm of every term in this sum.
So, in many cases (such as Hamiltonians with bounded strength and bounded range on finite-dimensional lattices), $JK$ and $\Vert H_{AB} \Vert$ will both be
proportional to the boundary area of $A$.

We now show Eq.~(\ref{bnd}).
Let
\be
\rho_0=\cZ_0^{-1} \exp[-\beta (H_A+H_B)],
\ee
where $\cZ_0$ is chosen so that ${\rm tr}(\rho_0)=1$.
We have
\be
\log(|\rho^{\Gamma_A}|_1) = \log(\frac{\cZ_0}{\cZ}) + \log(|\frac{\cZ}{\cZ_0} \rho^{\Gamma_A}|_1),
\ee
We bound the two terms on the right-hand side separately.

We will bound the first term by
\be
\label{freeenchange}
\log(\frac{\cZ_0}{\cZ}) \leq \beta \Vert H_{AB}\Vert.
\ee
This bound has a simple physical interpretation that the change in free energy between two Hamiltonians differing by adding the terms $H_{AB}$ is bounded
by the strength of the terms that we add.
To show this, we write a series:
\be
\label{series}
\frac{\cZ_0}{\cZ}={\rm tr}(\rho_0 \Bigl(1+\int_0^\beta {\rm d}\tau H_{AB}(\tau)+\int_0^{\beta} {\rm d}\tau_1 \int_0^{\tau_1} {\rm d}\tau_2 H_{AB}(\tau_1) H_{AB}(\tau_2) + \ldots\Bigr) ),
\ee
where we define $H_{AB}(\tau)$ as a time-evolved operator:
\be
O(\tau)\equiv \exp[\tau H] O \exp[-\tau H].
\ee
Consider the $n$-th term in the series and consider a given choice of $\tau_1\geq... \geq\tau_n$ in the integral for this term.  The integrand is the trace
${\rm tr}(\rho_0 H_{AB}(\tau_1) \ldots H_{AB}(\tau_n))$.
We write the expression inside the trace as a product of $2n$ operators, $n$ of which are operators $H_{AB}$ and the other $n$ of which are exponentials of $H$ :
\begin{eqnarray}
\label{t}
&& {\rm tr}(\rho_0 H_{AB}(\tau_1) \ldots H_{AB}(\tau_n)) \\ \nonumber
&=&
\cZ^{-1} {\rm tr}(H_{AB}\exp[-(\tau_1-\tau_2)H] H_{AB} \exp[-(\tau_2-\tau_3)H] \ldots \\ \nonumber && H_{AB} \exp[-(\tau_{n-1}-\tau_n) H]
H_{AB} \exp[-(\tau_n+\beta-\tau_1) H]).
\end{eqnarray}
We now use an inequality (Kristof's generalization\cite{k} of von Neumann's trace inequality)
that the trace (or trace norm) of a product of operators $O_1 O_2 ...$ with given eigenvalues is maximized in the case that all the operators are diagonal
and the eigenvalues are all in the same order (descending or ascending) along the diagonal.  Thus, if $H$ has eigenvalues $\lambda_i$ in ascending order ($\lambda_1 \leq \lambda_2 \leq ...$) and $H_{AB}$ has eigenvalues $\xi_i$ in descending order ($\xi_1 \geq \xi_2 \geq ...$) then
Eq.~(\ref{t}) is bounded by
\begin{eqnarray}
\label{t2}
&& {\rm tr}(H_{AB}\exp[-(\tau_1-\tau_2)H] H_{AB} \exp[-(\tau_2-\tau_3)H] \ldots \\ \nonumber && H_{AB} \exp[-(\tau_{n-1}-\tau_n)H]
H_{AB} \exp[-(\tau_n+\beta-\tau_1)H]) \\ \nonumber
& \leq & \sum_i \xi_i \exp[-(\tau_1-\tau_2) \lambda_i] \ldots \xi_i \exp[-(\tau_n+\beta-\tau_1) \lambda_i] \\ \nonumber
&=& \sum_i \xi_i^n \exp(-\beta \lambda_i) \\ \nonumber
&\leq & \Vert H_{AB} \Vert^n \sum_i \exp(-\beta \lambda_i) \\ \nonumber
&=& \Vert H_{AB} \Vert^n \cZ.
\end{eqnarray}
An alternative way to derive the bound in Eq.~(\ref{t2}) is to use H\"{o}lder's inequality\cite{Bhatia}: the trace norm of a product of $m$ matrices
is bounded by the product of the $p_i$ norms of the matrices: $|M_1 ... M_m|_1 \leq \prod_i |M_i|_{p_i}$ if $\sum_i p_i^{-1}=1$.
We take the $p=\infty$ norm (which is the operator norm) for the operators $H_{AB}$ in the product.  The other operators in the product
are all of the form $\exp(-\delta_i H)$ for some $\delta_i$; for those operators we take the $\beta/\delta_i$-norm which is equal to $\cZ_0^{\delta_i/\beta}$.  Since $\sum \delta_i=\beta$, indeed $\sum p_i^{-1}=1$ and $\prod_i |M_i|_{p_i}=\Vert H_{AB} \Vert^n \cZ$.

Hence, after integrating over $\tau_1,...,\tau_n$, the $n$-th term in the series is bounded by
$$\frac{(\beta \Vert H_{AB} \Vert)^n}{n!},$$ and so after summing over $n$ we get
\begin{eqnarray}
\label{Zratbnd}
\frac{\cZ_0}{\cZ} & \leq & \sum_{n \geq 0} \frac{(\beta \Vert H_{AB} \Vert)^n}{n!} \\ \nonumber
&=& \exp(\beta \Vert H_{AB} \Vert),
\end{eqnarray}
giving Eq.~(\ref{freeenchange}).

We now apply a similar series method to bound
$$\log(|\frac{\cZ}{\cZ_0} \rho^{\Gamma_A}|_1)=\log(|\cZ_0^{-1} \exp(-\beta H)^{\Gamma_A}|_1).$$
Expand $\cZ_0^{-1} \exp(-\beta H)$ as a series in $H_{AB}$:
\begin{eqnarray}
\cZ_0^{-1} \exp(-\beta H)&=&\cZ_0^{-1} \exp[-\beta (H_A+H_B+H_{AB})] \\ \nonumber
&=& \rho_0 \Bigl(1-\int_0^\beta {\rm d}\tau H_{AB}(\tau) + \ldots \Bigr),
\end{eqnarray}
where now we define $H_{AB}(\tau)$ as a time-evolved operator using Hamiltonian $H_A+H_B$ rather than $H$:
\be
O(\tau)\equiv \exp[\tau (H_A+H_B)] O \exp[-\tau (H_A+H_B)].
\ee
Using Eq.~(\ref{sumprod}),
expand the integrand in the $n$-th term in the series as a sum of products; for given $\tau_1 \geq ... \geq \tau_n$ this is a sum
$$\sum_{\alpha_1,...,\alpha_n} \rho_0 H_A^{\alpha_1}(\tau_1) H_B^{\alpha_1}(\tau_1) \ldots H_A^{\alpha_n}(\tau_n) H_B^{\alpha_n}(\tau_n).$$
The partial transposition operator is linear, so this gives us also a series for $\frac{\cZ}{\cZ_0} \rho^{\Gamma_A}$:
\begin{eqnarray}
&& \frac{\cZ}{\cZ_0} \rho^{\Gamma_A}\\ \nonumber &=& \rho_0^{\Gamma_A}+ \sum_{n\geq 1} (-1)^n \int_0^{\beta} {\rm d}\tau_1 \ldots \int_{0}^{\tau_{n-1}} {\rm d}\tau_n \sum_{\alpha_1,...,\alpha_n} \rho_0^{\Gamma_A} H_A^{\alpha_1}(\tau_1)^T H_B^{\alpha_1}(\tau_1) \ldots H_A^{\alpha_n}(\tau_n)^T H_B^{\alpha_n}(\tau_n).
\end{eqnarray}
By a triangle inequality,
\begin{eqnarray}
&& |\frac{\cZ}{\cZ_0} \rho^{\Gamma_A}|_1
\\ \nonumber &\leq &1+\sum_{n \geq 1} \int_0^{\beta} {\rm d}\tau_1 \ldots \int_0^{\tau_{n-1}} {\rm d}\tau_n
\sum_{\alpha_1,...,\alpha_n} \Bigl |\rho_0^{\Gamma_A} H_A^{\alpha_1}(\tau_1)^T H_B^{\alpha_1}(\tau_1) \ldots H_A^{\alpha_n}(\tau_n)^T H_B^{\alpha_n}(\tau_n) \Bigr|_1.
\end{eqnarray}

We can bound
$$|\rho_0^{\Gamma_A} H_A^{\alpha_1}(\tau_1)^T H_B^{\alpha_1}(\tau_1) \ldots H_A^{\alpha_n}(\tau_n)^T H_B^{\alpha_n}(\tau_n)|_1$$
as follows.  We have
\begin{eqnarray}
&& \rho_0^{\Gamma_A} H_A^{\alpha_1}(\tau_1)^T H_B^{\alpha_1}(\tau_1) \ldots H_A^{\alpha_n}(\tau_n)^T H_B^{\alpha_n}(\tau_n) \\ \nonumber
&=&
\Bigl( \exp(-\beta H_A^T) H_A^{\alpha_1}(\tau_1)^T \ldots H_A^{\alpha_n}(\tau_n)^T \Bigr)
\Bigl( \exp(-\beta H_B) H_B^{\alpha_1}(\tau_1) \ldots H_B^{\alpha_n}(\tau_n) \Bigr),
\end{eqnarray}
and so
\begin{eqnarray}
\label{fact}
&&|\rho_0^{\Gamma_A} H_A^{\alpha_1}(\tau_1)^T H_B^{\alpha_1}(\tau_1) \ldots H_A^{\alpha_n}(\tau_n)^T H_B^{\alpha_n}(\tau_n)|_1 \\ \nonumber
&=&
\cZ_0^{-1} \Bigl|\exp(-\beta H_A^T) H_A^{\alpha_1}(\tau_1)^T \ldots H_A^{\alpha_n}(\tau_n)^T\Bigr|_{1,A}
\Bigl|\exp(-\beta H_B) H_B^{\alpha_1}(\tau_1) \ldots H_B^{\alpha_n}(\tau_n)\Bigr|_{1,B},
\end{eqnarray}
where we introduce the notation that $|\ldots |_{1,A}$ means the trace norm of the given operator considered as an operator on the Hilbert space of system $A$, rather than on the full Hilbert space.  That is, we normalize the trace norm so that $|I|_{1,A}=d_A$ where $I$ is the identity operator on $A$ and $d_A$ is the dimension of the Hilbert space on $A$.  We define $|\ldots |_{1,B}$ similarly.

We can bound
$|\exp(-\beta H_A^T) H_A^{\alpha_1}(\tau_1)^T \ldots H_A^{\alpha_n}(\tau_n)^T|_{1,A}$ by writing this as a trace norm of a product of $2n$ operators:
$$\Bigl|(H_{A}^{\alpha_1})^T\exp[-(\tau_1-\tau_2)H_A^T]  \ldots (H_A^{\alpha_n})^T \exp[-(\tau_{n}+\beta-\tau_1)H_A^T]\Bigr|_{1,A}$$ and using
the same bound as was used to derive Eq.~(\ref{t2}), i.e. either Kristof's inequality or H\"{o}lder's inequality.
We have a bound $\Vert H_A^\alpha \Vert \Vert H_B^\alpha \Vert \leq J$; we can thus multiply $H_A^\alpha$ by a scalar and $H_B^\alpha$ by the inverse of that scalar so thatwe have the bound $\Vert H_A^\alpha \Vert, \Vert H_B^\alpha \Vert \leq \sqrt{J}$.
So, we bound the trace norm by
$|\exp(-\beta H_A^T)|_{1,A} J^{n/2}$.  We similarly bound
$|\exp(-\beta H_B) H_B^{\alpha_1}(\tau_1) \ldots H_B^{\alpha_n}(\tau_n)|_{1,B}$
So, the left-hand side of Eq.~(\ref{fact}) is bounded by $J^n$.
Summing over $\alpha_1,...,\alpha_n$, integrating over $\tau_1,...,\tau_n$, the $n$-th order term in the series for $\frac{\cZ}{\cZ_0} \rho^{\Gamma_A}$ is bounded in trace norm by
$(JK \beta)^n/n!$.
Summing over $n$ gives
\be
\frac{\cZ}{\cZ_0} \rho^{\Gamma_A} \leq \exp(JK\beta).
\ee
Combining with Eq.~(\ref{Zratbnd}) gives the claimed result.

\section{Relation to Entanglement Entropy}
Note that a bound on the log negativity is {\it not} implied by a bound on the mutual information for the von Neumann entropy.  Consider a pure state between $A,B$ of the form
\be
\psi=\sum_n A(n) |n\rangle_A \otimes |n\rangle_B,
\ee
for some orthonormal bases $|n\rangle_A,|n\rangle_B$.  For a pure state, the log negativity is given by the $1/2$ Renyi entropy, $2 \log(\sum_n |A(n)|)$.
Choose the Schmidt coefficients to decay as $A(n)=c/n$, for $1 \leq n \leq N$, with $c$ chosen as a normalization coefficient.
Since $\sum_n 1/n^2$ converges, $c$ converges to a constant for large $N$.
The von Neumann entropy and all Renyi entropies $S_\alpha$ for $\alpha> 1/2$ converge to a constant for large $N$, but $\sum_n A(n)$ diverges logarithmically in $N$ so the log negativity diverges as $\log(\log(N))$.

\end{widetext}


\begin{thebibliography}{2}
\bibitem{footnote1} In models where the ground state has zero negativity, such as the Heisenberg ferromagnet,
entanglement can develop at finite temperatures, so entanglement is not always a monotonic function of temperature.
See e.g. \cite{sbose}
\end{thebibliography}
\end{document}